\documentclass[twocolumn,amsmath,amssymb,secnumarabic, nobibnotes, aps, prd]{revtex4-2}

\usepackage{slashed}   
\usepackage{hyperref}
\usepackage{graphicx}

\begin{document}

\title{  Hadronic rescattering to solve helicity puzzle in $B^+\to p\bar p \pi^+(K^+)$ decays}

\author{I. Bediaga}
\affiliation{Centro Brasileiro de Pesquisas F\'isicas, 
22.290-180 Rio de Janeiro, RJ, Brazil}
\author{T. Frederico}
\affiliation{Instituto Tecnol\'ogico de
Aeron\'autica, 12.228-900 S\~ao Jos\'e dos Campos, SP,
Brazil.}
 \author{M. A. Shalchi}
\affiliation{Centro Brasileiro de Pesquisas F\'isicas, 
22.290-180 Rio de Janeiro, RJ, Brazil}
\author{P. C. Magalh\~aes}
\email[corresponding author: ]{pcmaga@unicamp.br}
\affiliation{Institute of Physics Gleb Wataghin, University of Campinas-UNICAMP, Campinas, São Paulo 13083-859, Brazil}
\date{\today}% It is always \today, today,

\begin{abstract}
We explore the contribution of hadronic final state interactions (FSI) to propose a production mechanism and interpret the puzzle on helicity angle distributions in $B^+\to p\bar p \pi^+$ and $B^+\to p\bar p K^+$ decays. Experimental results indicate opposite helicity angle $\theta_p$ distributions in those two channels with the difference presenting a remarkable linear dependence on $\cos \theta_p$. 
We assume the production mechanism is
driven by $B^+\to xy\, m^+ \to p\bar p m^+$, where   $m =\pi$ or $K$, and $xy$ represents favorable mesonic decay channels producing $p\bar p$. We develop a model that includes three-body final state interaction between the $p\,,~\bar p$ and $\pi^+$ or $K^+$ considering the dominance of elastic channels  $\pi^+p$ and $K^+\bar p$  interactions below 2\,GeV/c$^2$. Our three-body framework with  FSI  explains qualitatively the observed opposite behavior of the helicity distributions and the observed linearity.
%the helicity .  
%
\end{abstract}

%\keywords{B decays,  three-body channels, kaons, pion,}

\maketitle

\section{Introduction}

\begin{figure}[t!]
\begin{center}
\hspace*{-8mm}
    \includegraphics[width=8.7cm]{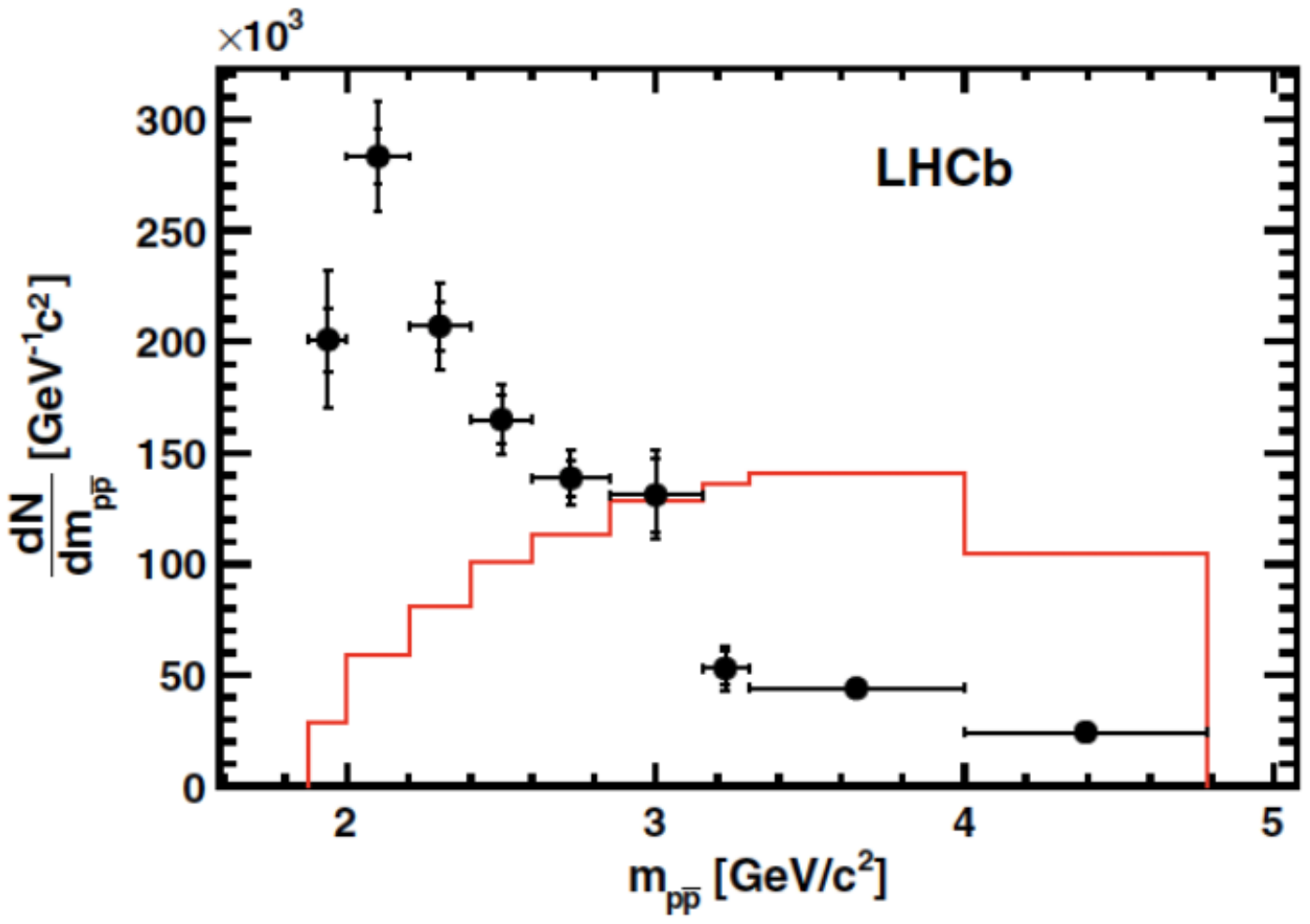}
   
    \vspace*{-10mm}
    \includegraphics[width=8.7cm]{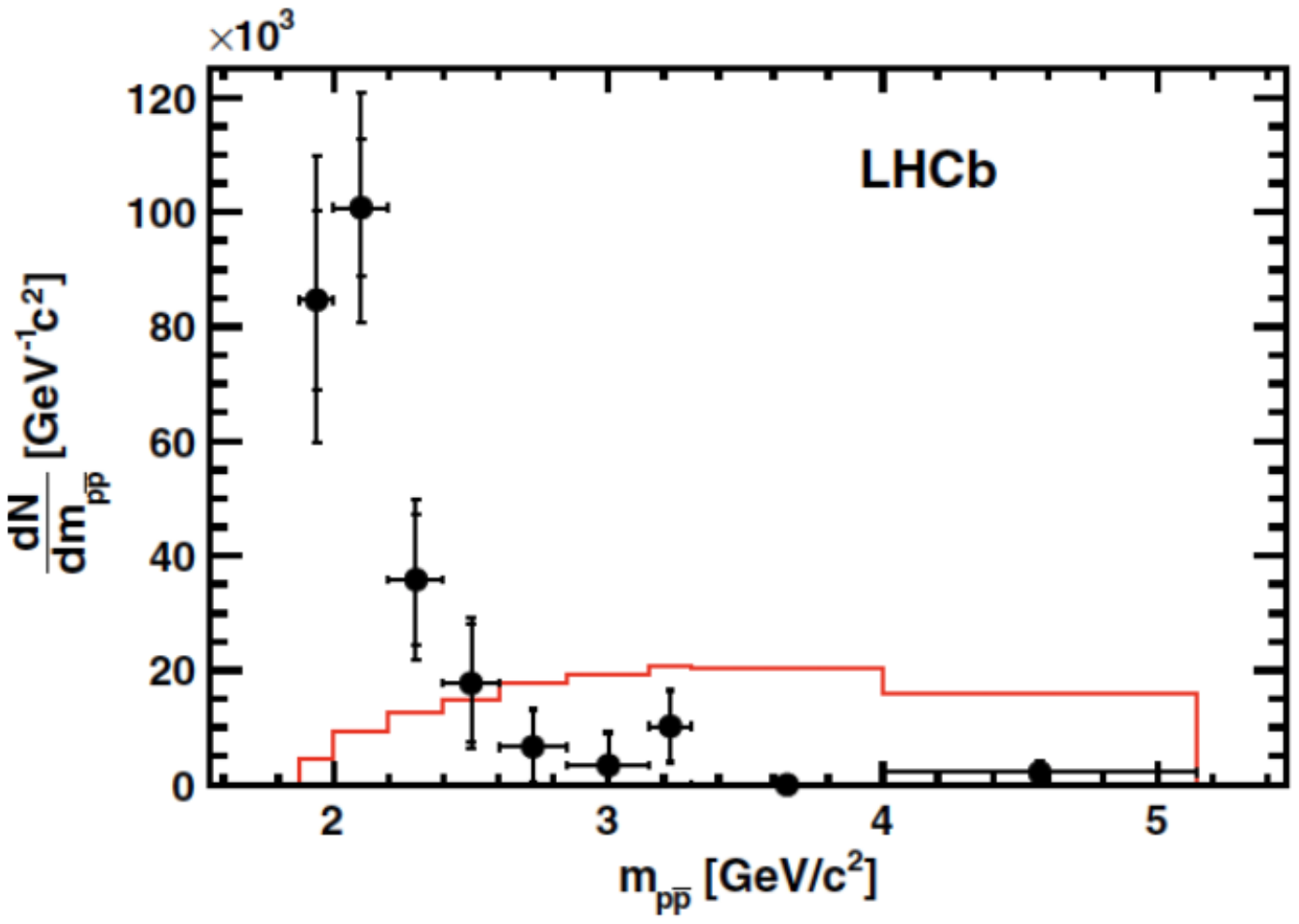}

  \includegraphics[width=8.7cm]{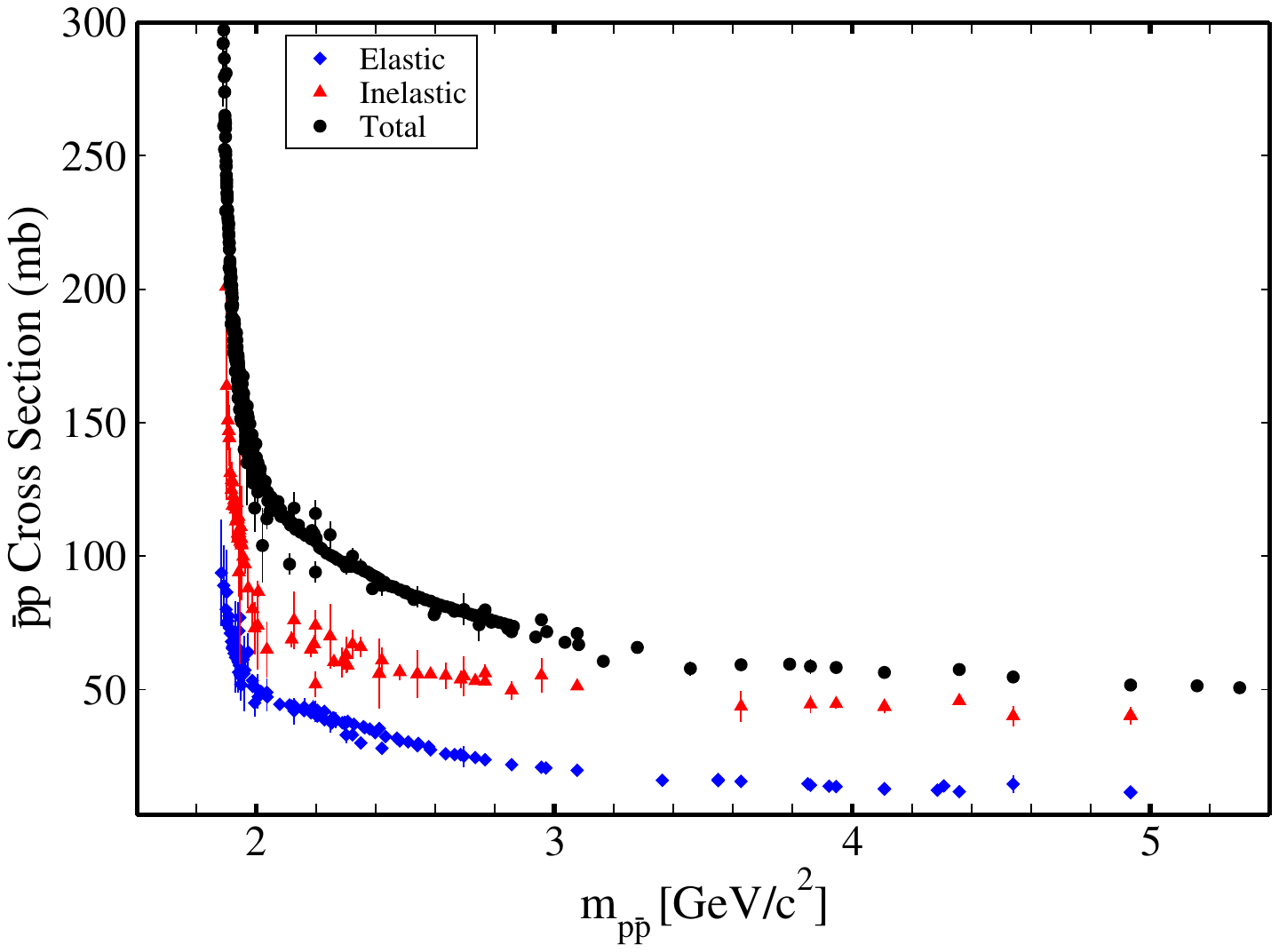}
    
       \caption{$p\bar p$ amplitude projections from (top) $B^+ \to p\bar p K^+$ decay and (middle) $B^+ \to p\bar p \pi^+$ decay, where the red line gives phase-space only contributions. Experimental values of the $p\bar p$  cross-sections from Ref.~\cite{PDG} (bottom): total
       (upper), inelastic (middle) and elastic (lower) cross-section data.  }
    \label{fig0}
    \end{center}
\end{figure}

Charmless three-body B-meson baryonic decays $B^+\to p\bar p \pi^+$ and $B^+\to p\bar p K^+$ called attention since the first experimental results~\cite{Belle:2002bro,BaBar:2005sdl,Belle:2007oni} due to two important features: i) the proton-antiproton invariant mass is placed near threshold and ii) the angular distribution for the $ B^+ \to p \bar p \pi^+$ decay is opposite to the $ B^+ \to p \bar p K^+$. Both features are confirmed and visible in more recent experimental results presented by LHCb collaboration~\cite{LHCb:2013njz,LHCbPRL2014}. Figure \ref{fig0} shown the LHCb ~\cite{LHCb:2013njz} $ p \bar p$ invariant mass distribution for $B^+\to p\bar p K^+$ and $B^+\to p\bar p \pi^+$ decays. The observed low mass proton-antiproton distributions is far from the expected pure phase-space population, indicating an important dynamical process taking place in the three-body final state.  

Similar $ p \bar p$ behavior was also observed in a large variety of decays, e.g.: $\text{J/}\Psi \to \gamma p \bar p$ \cite{BES:2003aic}, $B^0 \to \bar D^0 p \bar p$, $B^0 \to \bar D^{*0} p \bar p$, $B^0 \to D^-p \bar p \pi^+$\cite{BaBar:2011zka}, among other decays involving the proton-antiproton final state. To explain these non-trivial distributions, there are different phenomenological approaches:  threshold enhancement proposed by Soni and Hu \cite{Hou:2000bz} and the presence of new $p \bar p$ resonance associated with a glueball \cite{He:2005nm}. However, most of the phenomenological approach consider final state interaction (FSI) involving $ p \bar p$ re-scattering as the main dynamical mechanism responsible for this near-threshold behavior \cite{Haidenbauer:2006au,Kang:2015yka,DiSalvo:2015uoa,Laporta:2007zw}. 

In the present study involving charmless three-body $B$-meson baryonic decays $B^+\to p\bar p \pi^+$ and $B^+\to p\bar p K^+$ we assume that the re-scattering between $p$, $\bar p$ has an important role. The relevance of the re-scattering contribution in these final states is supported by three experimental features: (i) The $p\bar p$ low mass region is dominated by the inelastic scattering, see Fig.~\ref{fig0} (bottom), with most of the final states involving only light pseudo-scalar mesons final state that includes $\pi \pi $, $3\pi,\, 4\pi$, $K\bar K$ plus pions, and also scalar and vector meson resonances~\cite{Eastman:1973va,Chen:1977ik,Amsler:1997up}. (ii)  $B$ hadronic charmless  $B\to xy$ decays, where $x$ and $y$ can be a single or a combination of pseudo-scalar mesons like pions, kaons and eta, corresponding to the same mesonic final states observed in inelastic $p\bar p$ scattering, with a branching fractions up to three orders of magnitude larger than the ones for $B$ decays involving baryons in the final state~\cite{PDG}. 
And (iii) inspecting the $p\bar p$ mass distribution from the $B$-meson baryonic decays in Fig.~\ref{fig0} (top and middle) one can see they have a similar energy dependence to $p\bar p$ inelastic cross-section (bottom).
 
All together, these three pieces of experimental results, allow us to suppose that an important contribution from charmless $B$ mesonic decays is present in  $B-$ hadronic decays through $xy \to p\bar p$ re-scattering. In this work, we assume that charmless $B-$decay is the main source of $B^+\to p\bar p \pi^+$ and $B^+\to p\bar p K^+$ decays.
By assuming that, one can understand yet a fourth experimental feature, namely, the larger branching fraction of  $B$-meson-baryon decays, like $B^+\to p\bar p \pi^+ ( Br= 1.62\, \times 10^{-6})$ and $B^0\to p\bar p \pi^+\pi^- ( Br= 2.87 \times\,10^{-6})$ \cite{PDG}, compared to recent measured purely baryonic $B^0\to p \bar p \,(Br=1.25 \,\times 10^{-8})$  decay \cite{LHCbpbarp}. These decays
have the same weak vertex (short-distance contribution) but the latter is two orders of magnitude smaller. Following our proposal, the dominant mechanism in $B^0\to p \bar p$ decay
is through the light mesons re-scattering to $p\bar p$. However the production of $p \bar p $  is predominant below
%more important only around
2\,GeV, where the inelastic cross-section is large (see Fig.\,\ref{fig0}). Therefore near 5\,GeV, the energy of $B^0\to p \bar p$ decay,  $p\bar p$ inelastic cross section is strongly suppressed,  mainly its annihilation contribution \cite{Eastman:1973va}, giving a physical reason for the smaller branching fraction comparing to the other two decay channels. 

\begin{figure}[t!]
    \centering
    
\includegraphics[width=6.cm,angle=-90]{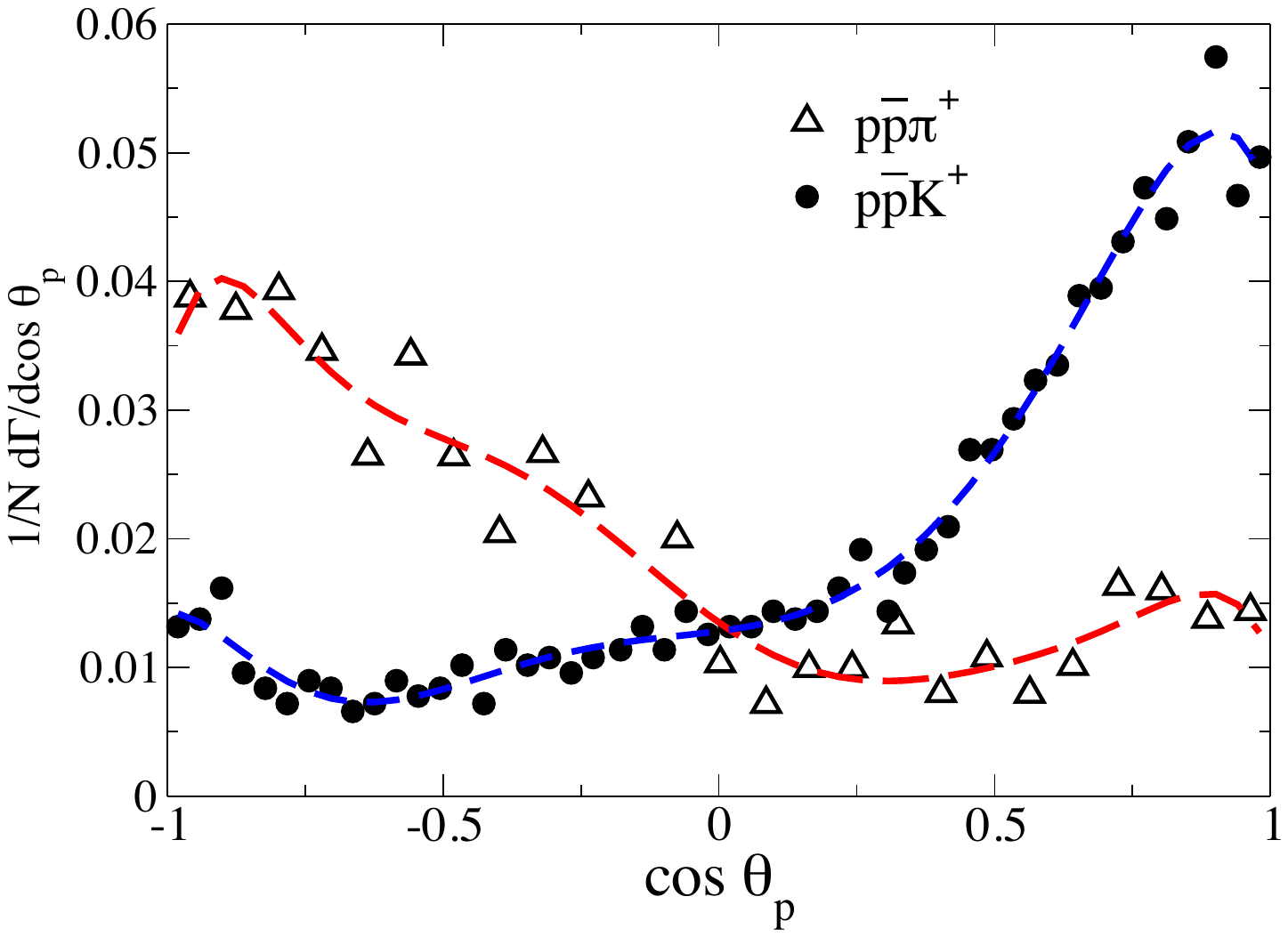}

\includegraphics[width=6.cm,angle=-90]{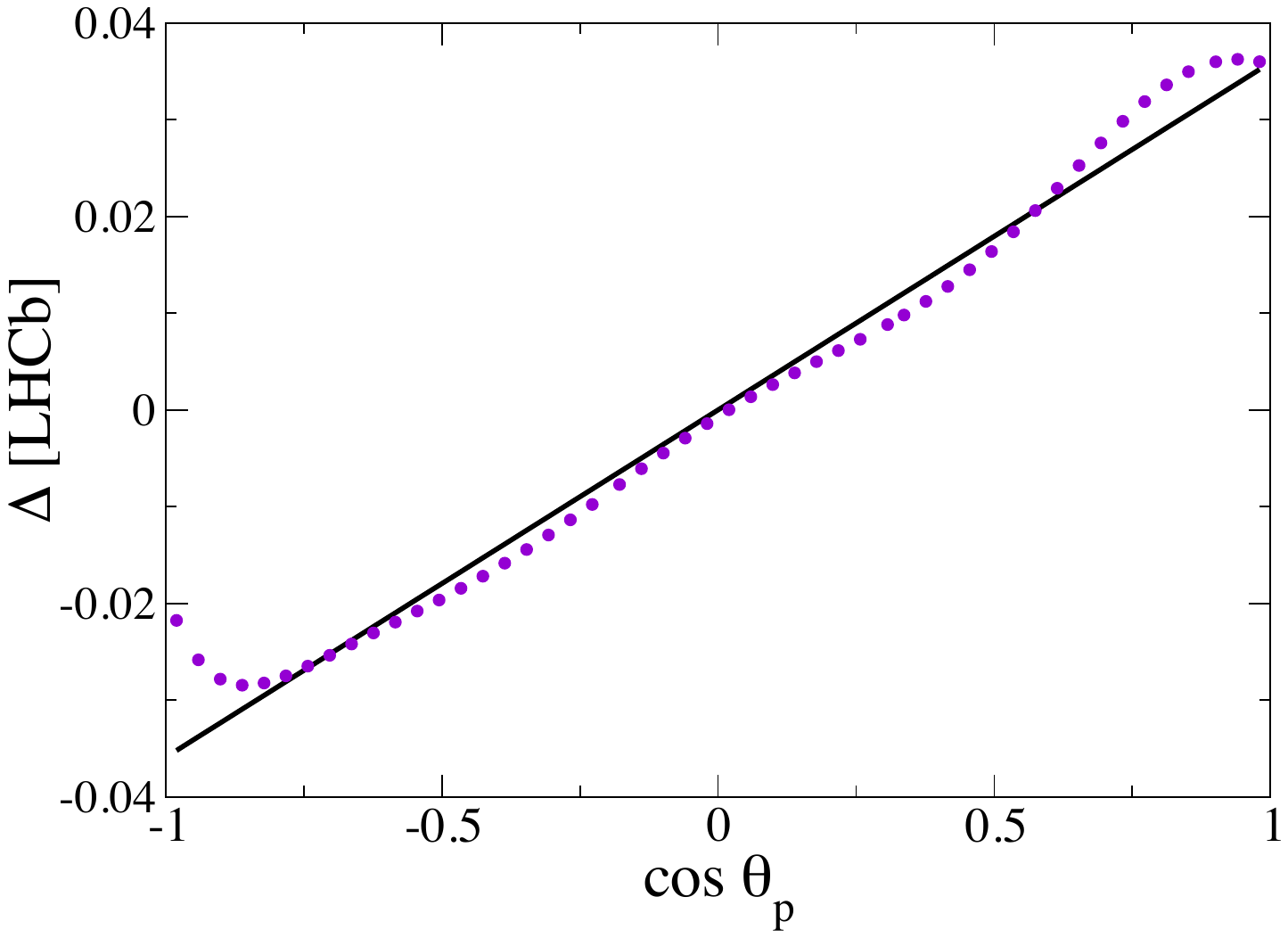}
    \caption{Top: LHCb data~\cite{LHCbPRL2014} for normalized distributions of $\cos\theta_p$ for the decay channels $B^+\to p\bar p\pi^+$ (triangles) and $B^+\to p\bar pK^+$ (circles), for integrated $M_{p\bar p}<\,$2.85\,Gev/c$^2$. For our qualitative analysis, the total uncertainties were not shown. The dashed lines are a polynomial fit $(\sum_{n=0}^8a_n\, \cos^n\theta_p)$ to the LHCb data. Bottom: Difference $\Delta^{B^+\text{\tiny {LHCb}}}_{p\bar pm}= \frac{1}{N}\frac{d\Gamma}{d\cos\theta_p}\big|_{p\bar pK^+}-\frac{1}{N}\frac{d\Gamma}{d\cos\theta_p}\big|_{p\bar p\pi^+}$ (dots) and the function $0.0359\cos\theta_p$ (solid line).}
    \label{fig2}
\end{figure}

 The second feature in $ B^+ \to p \bar p \pi^+$  and  $ B^+ \to p \bar p K^+$ decays is related to the helicity angle $(\theta_p)$, defined as the angle between the momentum of the charged meson and the oppositely charged baryon in the rest frame of the $p\bar p$ system. The LHCb data for the helicity angle distribution~\cite{LHCbPRL2014} is shown in the upper panel of Fig.~\ref{fig2}, where one notices that the angular dependence of the $ B^+ \to p \bar p \pi^+$ decay is opposite to the $ B^+ \to p \bar p K^+$ decay. In principle, there is no apparent reason to have a different helicity distribution if one doesn't consider a possible meson-baryon final state re-scattering in these decays. 

 The naive short-distance approach assumes that the particle associated with the heavy quark decay will carry more momentum than the one associated with the spectator quark~\cite{Cheng:2006bn,Suzuki:2006nn}. If this applies, then in the proton-antiproton rest frame both $B^+\to p\bar p \pi^+$ and $B^+\to p\bar p K^+$ helicity distribution must favor an antiproton more energetic, that is a negative helicity distribution. This naive mechanism agrees with the decay involving the pion in its final state and disagrees with the one with the kaon in the final state. 

 In the present paper, we take into account the possible final state interaction to explain both features observed in the experimental results, the low mass distribution observed in the proton-antiproton invariant mass and the different helicity distribution observed for these two charmless three-body B-meson baryonic decays.

In order to understand the difference between the two helicity distributions  in the $p\bar p \pi^+$ and $p\bar p K^+$ decay channels shown in the upper panel of Fig.~\ref{fig2}, we introduce the difference between them:
\begin{equation}
\Delta^{B^+\text{\tiny {LHCb}}}_{p\bar pm}= \frac{1}{N}\frac{d\Gamma}{d\cos\theta_p}\Big|_{p\bar pK^+}-\frac{1}{N}\frac{d\Gamma}{d\cos\theta_p}\Big|_{p\bar p\pi^+}\,.
\end{equation}
For the purpose of extracting the quantity $\Delta^{B^+\text{\tiny {LHCb}}}_{p\bar pm}$, we performed a polynomial fit to the LHCb data~\cite{LHCbPRL2014} using the function 
\begin{equation}
\frac{1}{N}\frac{d\Gamma}{d\cos\theta_p}\Big|_{p\bar pm}\approx\sum_{n=0}^8a_n\, \cos^n\theta_p
\end{equation}
for $m=\pi^+$ or $K^+$. The fit result is shown in the lines across the data in the upper panel of Fig.~\ref{fig2}, whereas the experimental $\Delta^{B^+\text{\tiny {LHCb}}}_{p\bar pm}$ is presented in the lower panel of the figure. Unexpectedly, $\Delta^{B^+\text{\tiny {LHCb}}}_{p\bar pm}$ presents a quite linear dependence on $\cos\theta_p$. 
Deviations are noticed for $|\cos\theta_p| \lesssim  0.75$ toward the kinematic boundary. This unexpected almost linear behavior of $\Delta^{B^+\text{\tiny {LHCb}}}_{p\bar pm}(\cos\theta_p)$ is not reported elsewhere in the literature. We aim to study this quantity with a decay amplitude that takes into account the hadronic FSI in a three-body re-scattering framework already applied to charmless $D$ decays~\cite{Magalhaes:2011sh,Guimaraes:2014kor}. 

In the present work, the three-body decay amplitudes 
for $B^+\to p\bar p\pi^+$  and $B^+\to p\bar pK^+$
are decomposed in the framework of the Faddeev-Bethe-Salpeter equations which includes the two-body scattering amplitudes $p\bar p$, $\pi^+p$, and $\pi^+\bar p$ 
in the first case.   For  $B^+\to p\bar pK^+$ 
the two-body interacting channels considered were $p\bar p$, $K^+p$, $K^+\bar p$. 
Such amplitude allowed us to perform a qualitative analysis to finally arrive at the linear dependence shown by the LHCb data and plotted in Fig.~\ref{fig2}. 

{
This work is organized as follows.  
Sect.~\ref{sec:decayamplitudefsi} presents the decay amplitude model with final state interaction. In this section,  the Faddeev-Bethe-Salpeter (FBS) formalism is reviewed, where the decomposition of decay amplitude in its Faddeev components 
can be computed from the driving or source amplitude. The contribution of the final state interaction to the decay amplitude model and the FBS equations for the bachelor amplitudes is also presented. This section is accompanied by Appendix A, where it is formally developed the FBS equations for the decay amplitude.  In Sect. ~\ref{sec:decayrate}, the $B^+\to p\bar p\pi^+$ and $B^+\to p\bar pK^+$ decay rates are explored, based on the FBS decomposition of the decay amplitude.
In Sect.~\ref{sec:qualitative} a qualitative analysis of the bachelor amplitudes is performed also based on the FBS equations, which is crucial for understanding the linear behavior of the difference in helicity distribution with 
$\cos\theta_p$. This discussion is accompanied by Appendix B. After that, the decay amplitude at low $M_{p\bar p}$ is formulated, which leads to an approximate formula for the helicity distribution asymmetry and the results are compared to the LHCb data. In Sect.~\ref{sec:summary} we provide the summary. Appendix C contains useful kinematic relations.
}

\section{Decay amplitude model with FSI}
\label{sec:decayamplitudefsi}
The full decay amplitude as (see \cite{Magalhaes:2011sh,Guimaraes:2014kor})
is described by a re-scattering series involving the final state $p\bar p\,m^+$ ($m=\pi, K$), and summed up in the $3\to
3$ transition matrix, $T_{3\to 3}$:
\begin{eqnarray}
{\cal A}(k_{p},k_{\bar p})=D(k_{p},k_{\bar p}
)
\\ +\int \frac{d^4q_p d^4q_{\bar p}}{(2\pi )^8}\,T_{3\to 3}(k_{p},k_{\bar p};q_{p},q_{\bar p})S_p(q_p)\,S_{\bar p}(q_{\bar p})\\
\times S_m(P-q_{p}-q_{\bar p})
D(q_p,q_{\bar p}) \ .\label{eq:Ampl}
\end{eqnarray}
Where the momentum of the proton and antiproton  are
$k_p$ and $k_{\bar p}$, and $P$ the total momentum. The bare off-shell  meson propagator is $S_{\pi,K}(k)=i(k^2-m_{\pi,K}^2+i \varepsilon)^{-1}\, ,$ and $S_{ p(\bar p)}$  is the Dirac propagator for proton/antiproton.
$D(k_p,k_{\bar p})$ is the driving term, which gives the production mechanism from the primary $B$ decay, i.e. source of the proton, antiproton, and meson in the final state. The operator $T_{3\to 3}$ is the T-matrix that accounts for the three-body re-scattering series in the  $p\bar pm$ system.

\begin{figure}[h!]
\begin{center}
\includegraphics[width=6cm,angle=0]{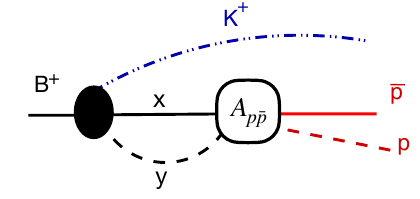}
\caption{Representation of the $B$ decay into a favored $B^+\to xyK^+$ channel and by re-scattering,  $xy\to p\bar p$, produces the final state channel. $xy$ is a short way to represent the favored mesonic channels with $K^+$ for the $B^+$ decay.}
\label{fig:Btohhbarktoppbark} 
\end{center}
\end{figure}
 As discussed in the previous section, the driving term of the decay amplitude for   $B^+\to p\bar p m^+$ in Eq.~\eqref{eq:Ampl}, is assumed to be dominated by meson decay channels $B^+\to xym^+$ that produce the $p\bar p$ pair by re-scattering of a neutral meson pair $xy\to p\bar p$, which is the source of the final
state. Diagrammatically, this process is represented in  Fig.~\ref{fig:Btohhbarktoppbark} for the particular case of $B^+\to p\bar p K^+$ decay. 

\begin{figure*}[t]
\begin{center}
\includegraphics[width=16cm,angle=0]{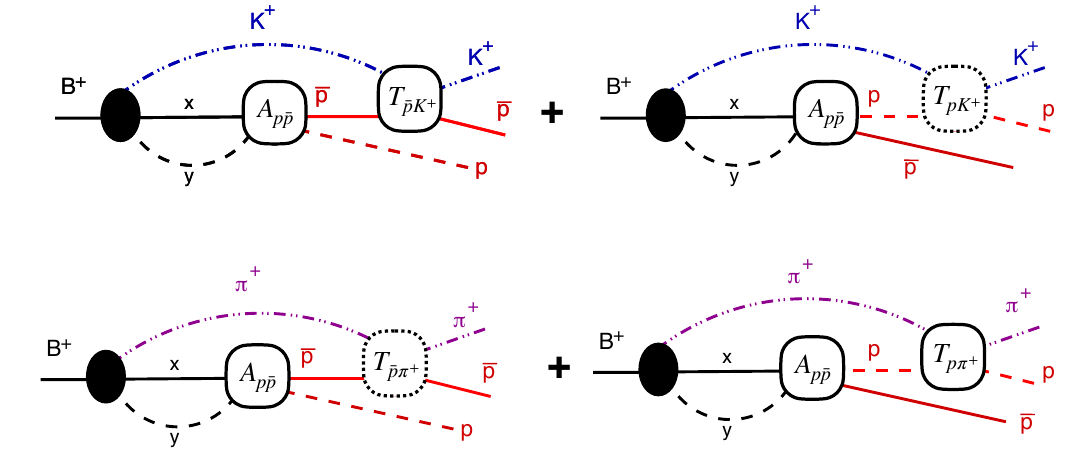}
\caption{Top panel: Diagrammatic representation of the contribution to the $B$ decay into a favored $B^+\to p\bar p K^+$ channel and by re-scattering,  $x\,y \to p\bar\, p$, produces the final state channel, including the elastic processes $K^+p\to K^+p$ and $K^+\bar p\to K^+\bar p$. Bottom panel: Diagrammatic representation of the contribution to the $B$ decay into a favored $B^+\to x\,y\,\pi^+$ channel and by re-scattering,  $x\,y\to p\bar p$, produces the final state channel, including the elastic processes $\pi^+p\to \pi^+p$ and $\pi^+\bar p\to \pi^+\bar p$.}
\label{fig:3bodyrescatt} 
\end{center}
\end{figure*}

The hadronic final state interactions (FSI) in Eq.~\eqref{eq:Ampl} are encoded in 
the $3\to 3$
transition matrix $T_{3\to 3}(k_{p},k_{\bar p};q_{p},q_{\bar p})$ convoluted with the driving term. 
In our framework, the
 FSI contribution is described by  
 connected diagrams from the scattering series corresponding to the ladder approximation. All possible  $2\to 2$ interaction terms are
summed up in the $3\to 3$ transition matrix. This represents the Faddeev components of the full transition matrix, i.e., the ones that have as driving term the two-body T-matrix  $Kp$, $K \bar p$,  $\pi p$ and $\pi \bar p$, namely, $T_{Kp}$, $T_{K\bar p}$, $T_{\pi p}$, $T_{\pi\bar p}$, as discussed in detail in Appendix~\ref{app:FBS}.   Fig.~\ref{fig:3bodyrescatt} illustrates the contribution of the first re-scattering terms to the decay amplitude and corresponds to the second and third terms of the right-hand side of Eq.~\eqref{eq44-1}. 

\subsection{Three-body re-scattering model}

A workable framework to explicitly build the Faddeev-Bethe-Salpeter equations (FBS)  was developed in Ref.~\cite{Guimaraes:2014kor} and summarized in Appendix~\ref{app:FBS}. 

 The two-body transition matrix elements in the FBS description dress the two-body scattering amplitude of particles $j$ and $k$, $\tau_{jk}$, with the presence of the third particle $i$:
\begin{eqnarray}
T_{jk}(k'_j,k'_k;k_j,k_k)=(2\pi)^4\tau_{jk}(s_{jk}) \,S^{-1}_i(k_i)\,\delta(k'_i-k_i)
\,.~~ \label{eq42-1}
\end{eqnarray}
With $S^{-1}_i(k_i)$ the propagator of the companion particle $i$, $s_{jk}=(k_j+k_k)^2$ the invariant  Mandelstam  variables and, the delta function imply energy-momentum conservation. 

 The scattering amplitude for the $\ell$ partial wave can be written in terms of the S-matrix as:
\begin{eqnarray}
\tau_{\ell,mp}\left(M^2_{mp}\right)=4\pi\,\frac{M_{mp}}{|\mathbf{k}^m_{mp}|}
\left(S_{\ell,mp}-1\right) \, ,\label{eq:tau12}
\end{eqnarray}
where $|\mathbf{k}^m_{mp}|$  is the modulus of the meson tri-momentum in the rest frame of the pair, $S_{\ell,mp}=\eta_{\ell} \text{e}^{i\,2\,\delta^\ell_{mp}}$, with $\eta_\ell $ the absorption parameter and $\delta^\ell_{mp}$ the phase-shift.
We assume that the driving term is dominant in the S-wave, and therefore in what follows it is assumed $\ell=0$ 
in Eq.~\eqref{eq:tau12} and for our discussions the spin of the proton/antiproton will be disregarded.

The full decay amplitude considering all S-wave pairwise interactions reduces to 
\begin{eqnarray}
{\mathcal A}(k_i,k_j)
=D(k_i,k_j)
+\sum_{a}\tau_{bc}(s_{bc})\,f_a(k_a) \ ,
\label{eq48}
\end{eqnarray}
where  $f_\alpha(k_\alpha)$ is the amplitude of the accompanying hadron and  carries the three-body re-scattering dynamics. In the present separable interaction model of Eq.~\eqref{eq42-1}, the Faddeev component of the three-body amplitude $ F_i(k_j,k_k)$, Eq.~\eqref{eq42}, is simplified to:
\begin{equation}
    F_i(k_j,k_k)= \tau_{jk}(s_{jk})f_i(k_i)\,. \label{eq:Di}
\end{equation}
and cyclic permutations of $\{i,j,k\}$.

 The connected FBS equations for the companion amplitudes $f_i(k_i)$ are derived by introducing the separable forms of the two-body T-matrix, Eq.~\eqref{eq42-1}, and the amplitudes~\eqref{eq:Di} in the Faddeev equations~\eqref{eq42}: 
\begin{multline}
f_i(k_i)=f_{0,i}(k_i)
\\+\int \frac{d^4q_j }{(2\pi )^4}S_j(q_j)S_k(P-k_i-q_j)\tau_j(s_{j})f_j(q_j) 
\\ +\int \frac{d^4q_k }{(2\pi )^4} S_j(P-k_i-q_k)S_k(q_k)\tau_k(s_{k})f_k(q_k) \,,
\label{eq47}
\end{multline}
where 
$f_{0,m}(k_m)= 0$
 to avoid double counting as the driving term with the production of $p\bar p$ already has the effect of the FSI. The non-vanishing first order %driving 
 terms are:
 \begin{equation} 
f_{0,p}(k_p)= \int \frac{d^4k_m }{(2\pi )^4}S_m(k_m)S_{\bar p}(k_{\bar p})D(k_p,k_{\bar p})\,, \label{eq47-D2}
\end{equation}
where ${k_{\bar p}=P-k_p-k_m}$, and for $f_{0\bar p}(k_{\bar p})$ one just change $k_p\leftrightarrow k_{\bar p}$ in the above definition.

As a matter of illustration, if one introduces the first-order terms given by Eq.~\eqref{eq47-D2} in the decay amplitude written in Eq.~\eqref{eq48} one gets the amplitudes corresponding to the diagrams represented in Fig.~\ref{fig:3bodyrescatt}.  
The full re-scattering series for the decay amplitude can be seen by iterating  Eqs.~\eqref{eq47} starting with the first-order terms.

\section{Decay rates }\label{sec:decayrate}

The decay rate for  $B^+\to p\bar p \pi^+$   and $B^+\to p\bar p K^+$ is obtained from the squared  modulus  of the decay amplitude  given in Eq.~\eqref{eq48}: 
\begin{multline}
   \Big| \mathcal{A}^{B^+}_{p\bar p m^+}\Big|^2= \Big| D^{B^+}_{ p\bar p m^+} 
+\tau_{p\bar p}(M^2_{p\bar p})\,f_{m} \big(|\mathbf{k}^{m}_{p\bar p}|\big)
\\ + \tau_{m\bar p }(M^2_{m\bar p})\, f_{p} \big(|\mathbf{k}^{p}_{m\bar p}|\big)+\tau_{m p }(M^2_{m p} )\, f_{ \bar p} \big( |\mathbf{k}^{\bar p}_{mp}|\big)\Big|^2
\,.
\label{eq:cp1-12}
\end{multline}
 The companion amplitudes should be evaluated on each particle mass-shell. As the companion amplitudes are scalars, the  interaction and the driving term are defined for S-wave, therefore the only remaining dependence will be on the modulus of the three-momentum $|\mathbf{k}^{a}_{bc}|$ of the companion particle $a$ in the pair $bc$ rest-frame.

The decay rate can be written as:
\begin{multline}
\big|\mathcal{A}^{B^+}_{p\bar p m^+}\big|^2\approx
   \Big| H^{B^+}_{p\bar pm}(M^2_{p\bar p},|\mathbf{k}^{m}_{p\bar p}|)\Big|^2
  \\  
+2 \text{Re}\Big\{\tau_{m\bar p }(M^2_{m\bar p})\, f_{p} \big(|\mathbf{k}^{p}_{m\bar p}|\big)\Big[H^{B^+}_{p\bar pm}(M^2_{p\bar p},|\mathbf{k}^{m}_{p\bar p}|)\Big]^*
 \Big\}
 \\
 \hspace{-.25cm}+2 \text{Re}\Big\{\tau_{m p }(M^2_{m p})\, f_{\bar p} \big(|\mathbf{k}^{\bar p}_{m p}|\big)\Big[H^{B^+}_{p\bar pm}(M^2_{p\bar p},|\mathbf{k}^{m}_{p\bar p}|)\Big]^*
 \Big\}
  \,, \label{eq:cp1-14}
\end{multline}
with
\begin{equation}
H^{B^+}_{p\bar pm}(M^2_{p\bar p},|\mathbf{k}^{m}_{p\bar p}|)=   D^{B^+}_{ p\bar p m^+} 
+\tau_{p\bar p}(M^2_{p\bar p})\,f_{m} \big(|\mathbf{k}^{m}_{p\bar p}|\big)
\,,
\end{equation}
 where we left only the terms that carry the driving amplitude.
 
\begin{figure}[t!]
\begin{center}
\includegraphics[width=0.95\linewidth]{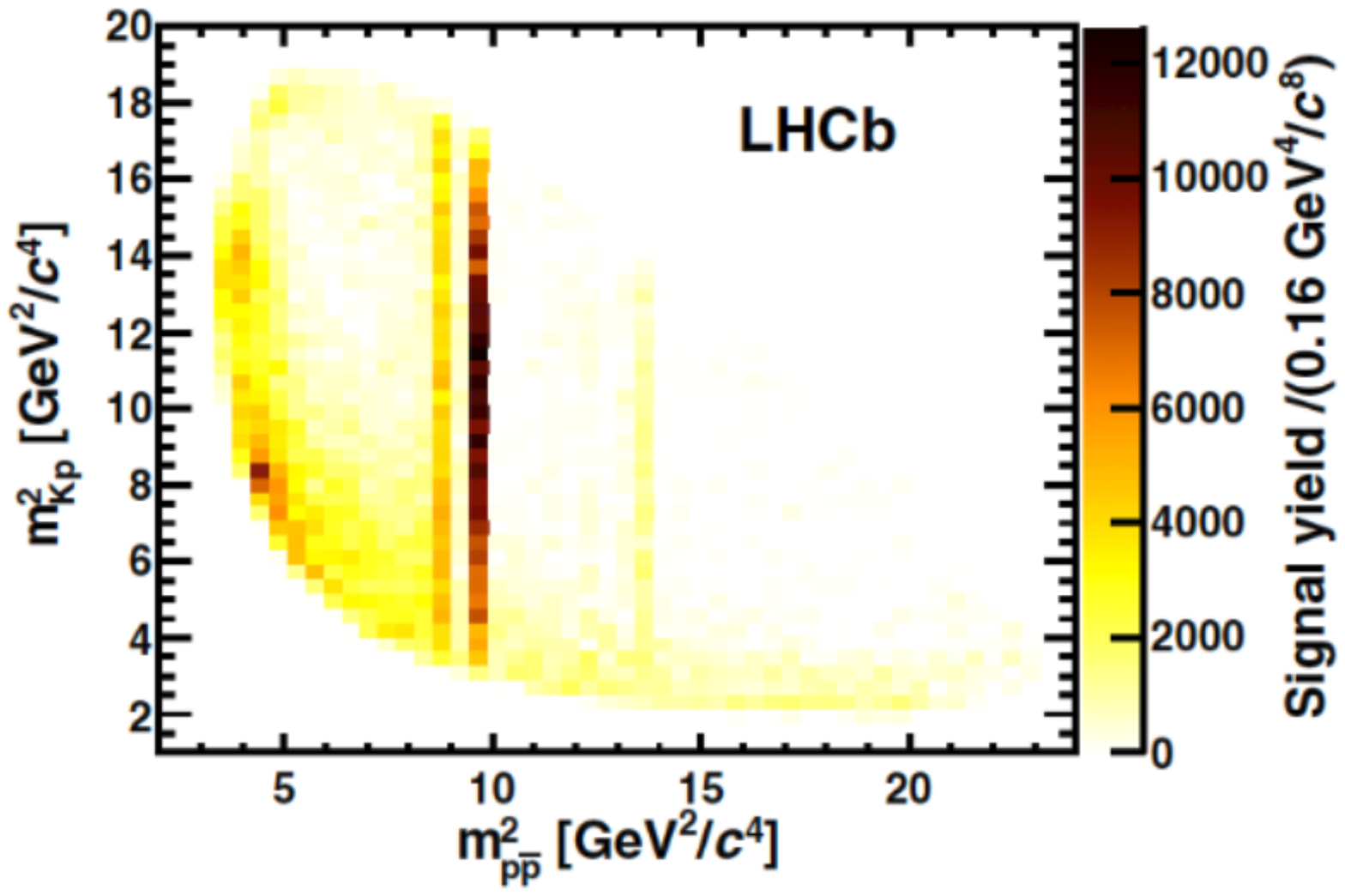}
\\
\includegraphics[width=0.95\linewidth]{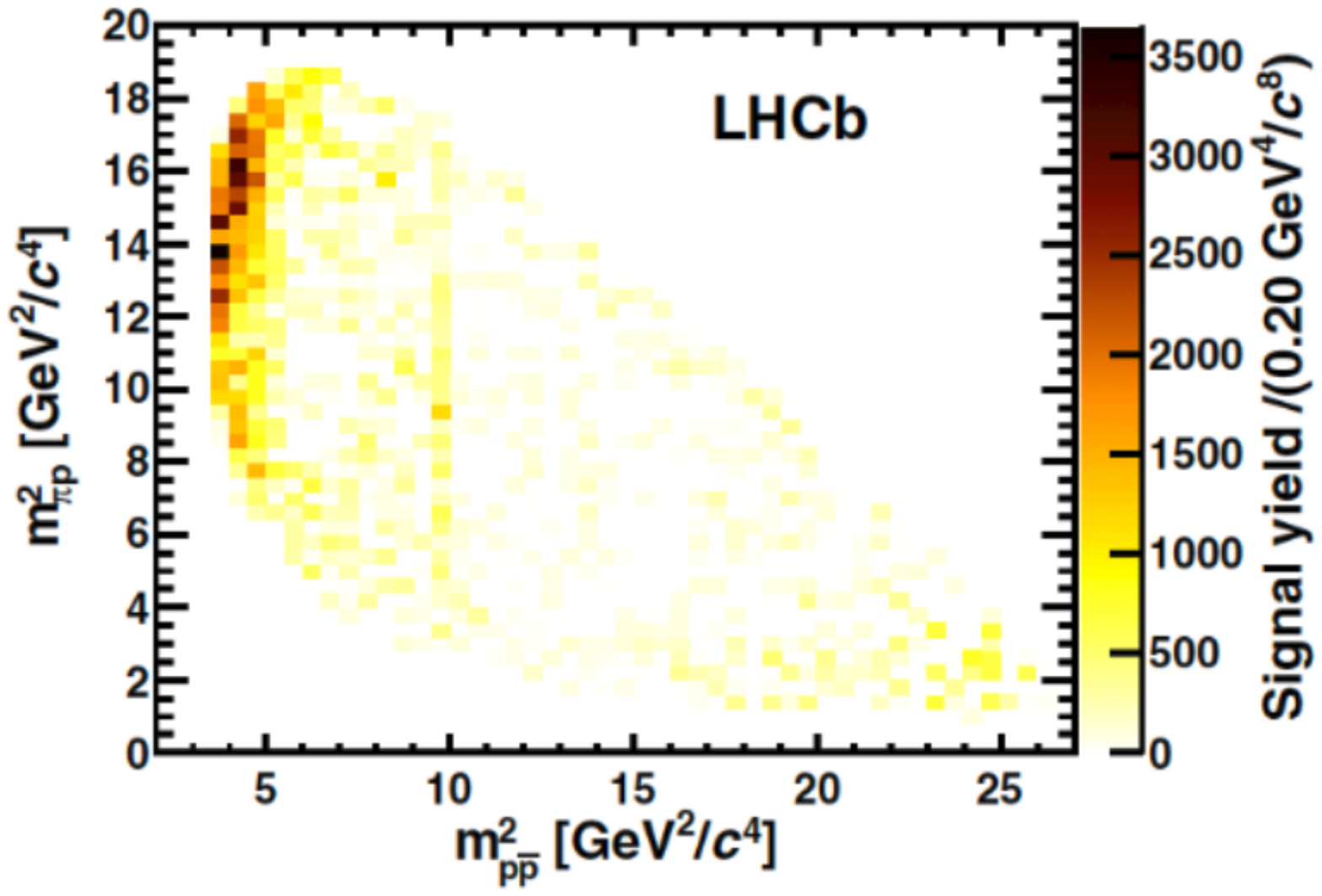}
\vspace{-.3cm}
\caption{Dalitz plots from Ref.~\cite{LHCbPRL2014} for $B^+\to p\bar p K^+$ (upper panel) and $B^+\to p\bar p \pi^+$ (lower panel) decays.  The ordinate axis has the mass of the neutral meson-antiproton pair. }
\label{fig1}
\end{center}
\end{figure}

 The expression of $B^+\to p\bar p m$ decay rate, Eq.~\eqref{eq:cp1-14}, highlights the FSI  role and bares the qualitative features found in the Dalitz distribution, shown in Fig.~\ref{fig1}, for low invariant masses of the $p\bar p$ pair.  
%
 %% Fig1
The vertical yellow bands in both $B^+\to p\bar p \pi^+$ and $B^+\to p\bar p K^+$ Dalitz plots in Fig.~\ref{fig1} for low $M^2_{p\bar p}$ comes from the large $p\bar p$ cross-section and the corresponding scattering amplitudes including the driving term from the threshold up to about 3-4~GeV$^2$/c$^4$.  
The lower yellow horizontal band in the $B^+\to p\bar p K^+$ Dalitz plot  (Fig.~\ref{fig1} up) is due to the relevant  $K^-p$ or $K^+\bar p$ cross section up to about 4\,GeV$^2$/c$^4$, while the  $K^+p$ or $K^-\bar p$ cross sections are significantly smaller than the neutral pair.
Curiously,  in $B^+\to p\bar p \pi^+$ Dalitz plot  (Fig.~\ref{fig1} low) the double charge $\pi^+ p$ cross-section is larger than the neutral pair. Indeed, one can barely see contribution from 
$\pi^+\bar p$ channel, whereas a strong interference at  $M^2_{\pi^+ p}$ between 12 and 18\,GeV$^2$/c$^4$ is observed. Kinematically, it corresponds to the low mass region of the $\pi^+p$ system, with large cross-section (see for reference~\cite{PDG2022}). 
In conclusion, the pattern seen in the Dalitz plots of the $B^\pm\to p\bar p \pi^\pm$ and $B^\pm\to p\bar p K^\pm$ decays can be attributed to the final state interaction between the pairs, which is encoded in Eq.~\eqref{eq:cp1-15}.

\section{Qualitative analysis}\label{sec:qualitative}

Our aim is to study the helicity angle distribution for the invariant mass of the $p\bar p$ system below 2.85\,GeV and where $M_{mp}$ and $M_{m\bar p}$ are large. This means that the momentum of the companion particle $|\mathbf{k}^{p}_{m\bar p}|$ and $|\mathbf{k}^{\bar p}_{mp}|$
are likely to be low. In order to obtain the decay amplitude,  $f_p(|\mathbf{k}^{p}_{m\bar p}|)$ and $f_{\bar p}(|\mathbf{k}^{p}_{mp}|)$ have to be  computed at low momenta and $f_{m} \big(|\mathbf{k}^{m}_{p\bar p}|\big)$ at high momenta. 

The dynamics that drives the companion functions $f_p$ and $f_{\bar p}$ at low momenta is associated with 
the  meson-baryon and $p\bar p$ interactions at low invariant masses. The latter is  more relevant at low mass (Fig.\,\ref{fig0}) and  interaction of $\pi^+p$ and $K^+\bar p$ ($K^-p$) are dominant over  $\pi^+\bar p$ ($\pi^-p$) and $K^+ p$, respectively (see cross-sections plots in Ref.~\cite{PDG2022}). 
 It is reasonable to expect that  the companion amplitudes, where the antiproton is spectator of the $\pi^+p$ system ($f_{\bar p,\pi^+}$), will behave similar to the one with the proton spectator of the $K^+\bar p$ system ($f_{ p,K^+}$), and analogously for the proton spectator of $\pi^+\bar p$ system ($f_{ p,\pi^+}$) and the antiproton spectator of $K^+p$ ($f_{ \bar p,K^+}$)\footnote{Note that  we introduce a new notation to avoid confusion.} 
The aforementioned similarity between the companion amplitudes that includes the FSI in the  $B^+\to p\bar p\pi^+$ and $B^+\to p\bar pK^+$ decays, we have:
\begin{equation}
\hspace{-.18cm} f_{p,\pi^+}\approx f_{\bar p,K^+}=f_{ p,K^-}~\text{and}~ f_{\bar p,\pi^+}\approx f_{p,K^+}=f_{\bar p,K^-}\,,
\label{eq:fkfpi}
\end{equation}
once it is assumed that the driving term $f_{0,p}=f_{0,\bar p}$, Eq.~\eqref{eq47-D2},
is symmetric by exchange of the $p\leftrightarrow\bar p$ and weakly dependent on the exchange of $\pi\leftrightarrow K$, considering that the amplitude, $D(k_i,k_j)$ holds these properties.

The qualitative argument that leads to Eq.~\eqref{eq:fkfpi} is further elaborated by analyzing the 
FBS equations given in  Appendix~\ref{app:fi} for the companion amplitudes, where we show that the physical role played by the proton in the $B^+\to p\bar p \pi^+$ is played by the antiproton in the $B^+\to p\bar p K^+$ decay.

Taking into account  the relations expressed in Eq.~\eqref{eq:fkfpi}  the useful companion amplitudes $f_{\pm}$ are introduced by:
\begin{eqnarray}
f_{-} &=&  \frac{ f_{p,\pi^+}-f_{\bar p,\pi^+}}{2}\approx \frac{f_{\bar p,K^+}- f_{p,K^+}}{2}\,,
    \label{eqasympik}
\\
    f_+&=&\frac{f_{p,\pi^+}+f_{\bar p,\pi^+}}{2}\approx\frac{f_{p,K^+}+f_{\bar p,K^+}}{2}\,,  \label{eqsympik}
\end{eqnarray}
%which will be relevant for the following analysis of the $B^+$ decay in these two channels. 
In particular, note that $f_-$ is antisymmetric by the exchange of $p\leftrightarrow \bar p$, a crucial property to understand the experimental observation of  the 
opposite behaviour of the helicity distribution for the $p p\bar\pi^+$ and $p\bar pK^+$ as seen in Fig.~\ref{fig2}, and also the
linear behavior of $\Delta^{B^+\text{\tiny {LHCb}}}_{p\bar pm}$ with $\cos\theta_p$, as it will be shown in what follows.

\subsection{Decay amplitude for $M_{p\bar p}<2.85\,$GeV}

The invariant mass of the $p\bar p$ system below 2.85\,GeV implies high  momentum collision in the meson-proton/antiproton above the resonance region, where the absorption to inelastic channels is supposedly to be large  and then $\eta\approx 0$ in the scattering amplitude~\eqref{eq:tau12}, which leads to:
\begin{equation}
 \hspace{-.28cm}   \tau_{m\bar p }(M^2_{m\bar p})\approx -\frac{4\pi M_{m\bar p}}{|\mathbf{k}^m_{m\bar p}|}~ \text{and}~ \tau_{m p }(M^2_{m p})\approx -\frac{ 4\pi M_{m p}}{|\mathbf{k}^m_{m p}|}.
\end{equation}
Then, Eq.\eqref{eq:cp1-14} is simplified to:
\begin{multline}
   \big|\mathcal{A}^{B^+}_{p\bar p m^+}\big|^2\approx
   \Big| H^{B^+}_{p\bar pm}(M^2_{p\bar p},|\mathbf{k}^{m}_{p\bar p}|)\Big|^2
  \\  
 -\frac{8\pi M_{m\bar p}}{|\mathbf{k}^m_{m\bar p}|} \text{Re}\Big\{f_{p} \big(|\mathbf{k}^{p}_{m\bar p}|\big)\Big[H^{B^+}_{p\bar pm}(M^2_{p\bar p},|\mathbf{k}^{m}_{p\bar p}|)\Big]^*
 \Big\}
 \\
 - \frac{8\pi M_{m p}}{|\mathbf{k}^m_{mp}|}\text{Re}\Big\{ f_{\bar p} \big(|\mathbf{k}^{\bar p}_{m p}|\big)\Big[H^{B^+}_{p\bar pm}(M^2_{p\bar p},|\mathbf{k}^{m}_{p\bar p}|)\Big]^*
 \Big\}
  \,. \label{eq:cp1-15}
\end{multline}

The decay amplitude of Eq.~\eqref{eq:cp1-15}  suggests a double peak for high invariant mass of both ${mp}$ and ${m\bar p}$ systems. That would lead to a symmetric dependence in $\cos\theta$, namely the helicity angle $\theta$ of the charged meson and the oppositely charged baryon in the rest frame of the $p\bar p$ system. However, the re-scattering that gives contributions beyond the inhomogeneous terms, $f_{0,p}$ and $f_{0,\bar p}$, should break the symmetry.

The relations written in Eq.~\eqref{eq:fkfpi}  express that the physical role played by the proton in the $B^+\to p\bar p \pi^+$ is played by the antiproton in the $B^+\to p\bar p K^+$ decay.
In order to make evident this property the functions $f_\pm$ defined in Eqs.~\eqref{eqasympik} and \eqref{eqsympik}
are introduced in Eq.~\eqref{eq:cp1-15} for the decay amplitude, which gives:  
\begin{eqnarray}
  &&\big|\mathcal{A}^{B^+}_{p\bar p m}\big|^2_\pm\approx
   \Big| H^{B^+}_{p\bar pm}(M^2_{p\bar p},|\mathbf{k}^{m}_{p\bar p}|)\Big|^2 \nonumber
  \\  
 &-&\frac{8\pi M_{m\bar p}}{|\mathbf{k}^m_{m\bar p}|} \text{Re}\Big\{\Big( f_+ \big(|\mathbf{k}^p_{m\bar p}|\big)\pm f_- \big(|\mathbf{k}^p_{m\bar p}|\big)\Big)\Big[H^{B^+}_{p\bar pm}\Big]^*
 \Big\}\quad \nonumber
 \\
 &-& \frac{8\pi M_{m p}}{|\mathbf{k}^m_{mp}|}\text{Re}\Big\{  \Big(f_+ \big( |\mathbf{k}^{\bar p}_{m p}|\big)
 \mp f_- \big( |\mathbf{k}^{\bar p}_{m p}|\big)\Big)\Big[H^{B^+}_{p\bar pm}\Big]^*
 \Big\}
  \,,\quad\quad \label{eq:cp1-15a}
\end{eqnarray}
where $ \big|\mathcal{A}^{B^+}_{p\bar p m}\big|^2_+$ stands for $p\bar p\pi^+$ and $ \big|\mathcal{A}^{B^+}_{p\bar p m}\big|^2_-$  for $p\bar pK^+$.

The difference of the decay rate for the two channels is given by:
\begin{eqnarray}
 \big|\mathcal{A}^{B^+}_{p\bar p \pi^+}\big|^2&-&\big|\mathcal{A}^{B^+}_{p\bar p K^+}\big|^2
  =  \nonumber \\ 
 &-&\frac{16\pi\,M_{m\bar p}}{|\mathbf{k}^m_{m\bar p}|}
 \,\text{Re}\Big\{f_- \big(|\mathbf{k}^p_{m\bar p}|\big)\Big[H^{B^+}_{p\bar pm}\Big]^*\Big\}
\nonumber \\
 &+&\frac{16\pi\,M_{m p}}{|\mathbf{k}^m_{mp}|} \text{Re}\Big\{f_- \big(|\mathbf{k}^{\bar p}_{mp}|\big)\Big[H^{B^+}_{p\bar pm}\Big]^*\Big\}
  \,,\label{eq:cp1-16}
\end{eqnarray} 
which indicates that the opposite behaviour observed  in high $M_{\pi^+\bar p}$ and low  $M_{K^+\bar p}$  Dalitz plot distribution, 
Fig.~\ref{fig1}, 
 reflects the contribution of the hadronic final state interaction in the $B^+$ decay. The mass difference between the pion and the kaon is disregarded in the kinematic region of $M_{p\bar p}<$2.85\,GeV/c$^2$.

Assuming a monotonic dependence of $f_\pm(|\mathbf{k}|)$ and $H^{B^+}_{p\bar pm}$ on $\mathbf{k}$ in the momentum region of interest, we can approximately write that:
\begin{equation}
   \big|\mathcal{A}^{B^+}_{p\bar p \pi^+}\big|^2-\big|\mathcal{A}^{B^+}_{p\bar p K^+}\big|^2\approx \zeta \Bigg(\frac{M_{m\bar p}}{|\mathbf{k}^m_{m\bar p}|}-\frac{M_{m p}}{|\mathbf{k}^m_{mp}|}  \Bigg)\, , \label{eqdiference}
\end{equation}
where $\zeta$ appears as an average value and  the LHCb data suggests $\zeta>0$ (see  Fig.~\ref{fig2}).
Furthermore, Eq.~\eqref{eqdiference} implies in an anti-symmetric dependence on  $\cos\theta_p$ in the helicity distribution, as we are going to detail in the following.

\subsection{Helicity distributions }
The  Eq.~\eqref{eqdiference} can be further elaborated using the invariant mass of the  $ mp$ pair as a function of  $\theta_p$,
\begin{equation}
\begin{aligned}
M^2_{ mp (m\bar p)} (\cos \theta_{p(\bar p)}) &=\frac12\Big(M^2_B-M_{p\bar p}^2+m^2_m +2m^2_{p} \Big)
\\ &+(-)\frac12\cos \theta_p \;\sqrt{M^2_{p\bar p}-4 m_p^2}
\\ &\times \sqrt{\Bigg({M^2_B-M_{p\bar p}^2-m^2_m \over \, M_{p\bar p}}\Bigg)^2-4m^2_m}
\,, 
    \end{aligned}
\end{equation}
and the squared momentum for the meson in the rest-frame of the $mp$ pair is:
\begin{equation}
    \begin{aligned}
&|\mathbf{k}^m_{mp}|^2=\Bigg({M^2_{mp}+m^2_m-m^2_p \over 2M_{mp}}\Bigg)^2 -\,m^2_m \, .
    \end{aligned}
    \label{eq:kmppbar-1}
\end{equation}
Note the opposite sign multiplying $\cos \theta_p$ in the invariant mass of the pairs $mp$ and $m\bar p$, which implies that the difference $ \big|\mathcal{A}^{B^+}_{p\bar p \pi^+}\big|^2-\big|\mathcal{A}^{B^+}_{p\bar p K^+}\big|^2$ given in Eq.~\eqref{eqdiference} is an odd function of $\cos\theta_p$.

Now an approximation is done by disregarding the mass of the meson and thus:
\begin{equation}
\begin{aligned}
M^2_{ mp} (\cos \theta_p)&=\frac12\Big(M^2_B-M_{p\bar p}^2 +2m^2_{p} \Big)
\\ &+\frac12\cos \theta_p \;\sqrt{1-4 \frac{m_p^2}{M^2_{p\bar p}}}\,\Big(M^2_B-M_{p\bar p}^2\Big)
\,, 
    \end{aligned}
\end{equation}
and
\begin{equation}
    \begin{aligned}
        &|\mathbf{k}^m_{mp}|^2={\Big(M^2_{pm}-m^2_p \Big)^2\over 4M^2_{pm}} \, ,
    \end{aligned}
    \label{eq:kmppbar-2}
\end{equation}

The ratio relevant for the difference in the decay rates becomes:
\begin{equation}
    \frac{M_{mp}}{|\mathbf{k}^m_{mp}|}=\frac{2\,M^2_{mp}}{M^2_{mp}-m^2_p}
\end{equation}
and the analogous expressions for ${M_{m\bar p}}/{|\mathbf{k}^m_{m\bar p}|}$, with that
\begin{equation} \big|\mathcal{A}^{B^+}_{p\bar p \pi^+}\big|^2-\big|\mathcal{A}^{B^+}_{p\bar p K^+}\big|^2=
\zeta\, \cos\theta_p\,\mathcal{F}(\cos^2\theta_p,M^2_{p\bar p})  \,,
\end{equation}
where the even function in $\cos\theta_p$ is
\begin{multline}
    \mathcal{F}(\cos^2\theta_p,M^2_{p\bar p})= \\= \frac{2\,m_p^2\,({M^2_B-M_{p\bar p}^2 })\,\sqrt{M^2_{p\bar p}-4 m_p^2}}{\big(M^2_{mp}-m^2_p\big) \,M_{p\bar p}\,\big(M^2_{m\bar p}-m^2_p\big)}
\, .
\end{multline}
The integral over the kinematic region of $M_{p\bar p}<\bar M_{p\bar p}=2.85\,$GeV/c$^2$ is:
\begin{equation}
     \Delta^{B^+}_{p\bar pm}=\zeta\,\cos\theta_p \, \int_{4m^2_p}^{\bar M^2_{p\bar p}} dM^2_{p\bar p}\,\,\mathcal{F}(\cos^2\theta_p,M^2_{p\bar p})\,,
     \label{eq:delta_1}
\end{equation}
that gives the difference in the helicity distributions in the $B^+\to p\bar p\pi^+$ and  $B^+\to p\bar pK^+$ decays.

\subsection{Results for  $ \Delta^{B^+}_{p\bar pm}$ with $M_{p\bar p}<2.85\,$GeV/c$^2$}

\begin{figure}
 \centering
 \includegraphics[width=8.5cm]{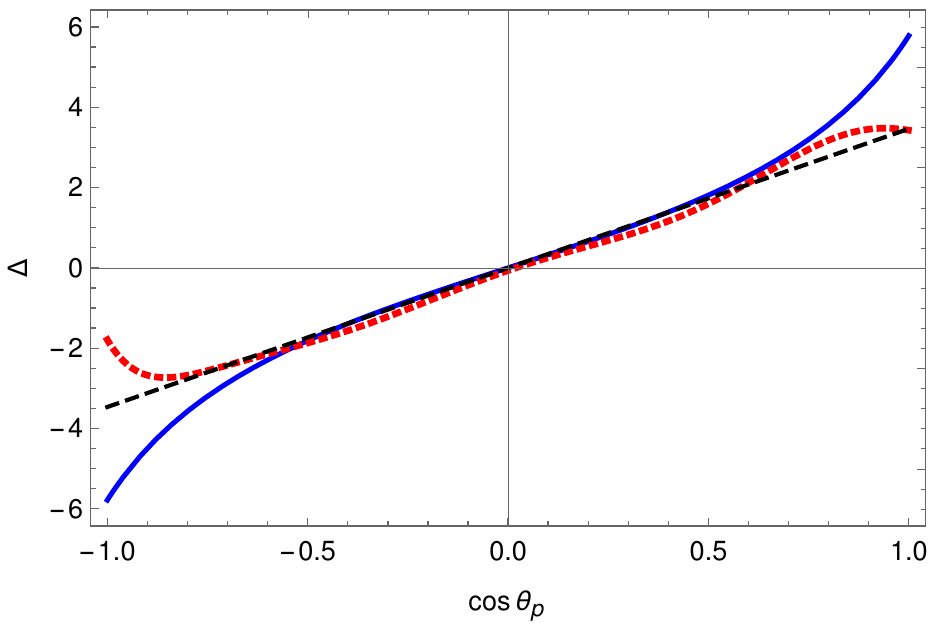}
    \caption{Comparison of the normalized $ \Delta^{B^+}_{p\bar pm}$  (solid line) with an estimated  $ \Delta^{B^+\text{\tiny LHCb}}_{p\bar pm}$  from the LHCb experimental data~\cite{LHCbPRL2014} (dotted line). The dashed line is the normalized $\cos\theta_p$. Each curve is normalized according to Eq.~\eqref{eq:normdelta}.  } 
    \label{fig:crosskpi}
\end{figure}

The results of $ \Delta^{B^+}_{p\bar pm}$ and the estimated $\Delta^{B^+\text{\tiny {LHCb}}}_{p\bar pm}$ are shown in Fig.~\ref{fig:crosskpi} using the normalization:
\begin{equation}
    ||\Delta^{B^+}_{p\bar pm}||=\Bigg[\int^\frac12_{-\frac12} d\cos\theta_p \Big(\Delta^{B^+}_{p\bar pm}(\cos\theta_p)\Big)^2\Bigg]^\frac12=1\,. \label{eq:normdelta}
\end{equation}

The approximate formula, Eq.~\eqref{eq:delta_1}, follows the linear behaviour of the data between  $-0.7\lesssim\cos\theta_p\lesssim 0.8$ while towards the kinematic end-points, there are discrepancies. The dominant linear dependence on $\cos\theta_p$ 
is the main result of our qualitative analysis with FSI contribution to $B^+\to p\bar p \pi^+$ and $B^+\to p\bar p K^+$ decays, based on the prevalence of the $\pi^+-p$ 
cross-section over the $\pi^--p\equiv\pi^+-\bar p$, and $K^+-\bar p$ over $K^+- p$ cross section below 2\,GeV/c$^2$ invariant mass.

For $M_{p\bar p}=2.85\,$GeV/c$^2$ the smallest invariant mass of the $\pi^+$-proton system,  1.83\,GeV/c$^2$, is placed close to $\cos\theta_p=- 1$   whereas for  $K^+\bar p$ subsystem, 1.9\,GeV/c$^2$, is placed at $\cos\theta_p=1$. The smaller masses attained in the integration over $M^2_{p\bar p}$ to compute $\Delta^{B^+}_{p\bar pm}$ shows the limitation of our model when approaching the resonance region, which is clearly exhibited in the comparison with the LHCb data towards  $|\cos\theta_p|\to 1$, that is even more evident for $\cos\theta_p=-1$ where the $\pi^+-$proton mass is smaller. Furthermore,  the approximation that leads to Eq.~\eqref{eqdiference} disregarded the momentum dependence of the companion amplitudes, $f_-(|\mathbf{k}|)$, in Eq.~\eqref{eq:cp1-16}, which should present a monotonic decrease with $|\mathbf{k}|$. Towards the kinematic end-points $|\mathbf{k}|$ is larger for $M_{p\bar p}<2.85\,$GeV/c$^2$, which could eventually play a role in the appreciable damping of the $ \Delta^{B^+}_{p\bar pm}$ magnitude away from the LHCb data.

\section{Summary}\label{sec:summary}

In this work, the contribution of the hadronic final state interaction
to the $B^+\to p\bar p\pi^+$ and $B^+\to p\bar pK^+$ decays is considered in a Three-body Faddeev-Bethe-Salpeter framework. 

In the FBS formulation, the source, or driving term, is assumed to be dominated by B-mesonic decays, with the physical reasons supporting it discussed in Sec. I. 

The two body 
re-scattering interactions used in FBS framework include
$\pi p$, $\pi\bar p$ and $p\bar p$   for $B^+\to p\bar p\pi^+$ decay, and $K p$, $K\bar p$ and $p\bar p$ amplitudes for $B^+\to p\bar pK^+$ decay, where it is assumed the dominance of s-wave scattering. These decay amplitudes present the qualitative features of the Dalitz distribution, namely the enhancement of the distribution at low
$p\bar p$ invariant masses and low $\pi^+p$ invariant mass in one case and low $K^+\bar p$ invariant mass in the other one.

Furthermore, it was qualitatively explored the decay amplitude model to interpret the asymmetry distribution of the helicity angle, $\theta_p$, 
as observed by the LHCb collaboration~\cite{LHCbPRL2014}. 

The experimental data reveals that the difference between the normalized helicity distributions for the invariant mass $M_{p\bar p}<2.85\,$GeV/c$^2$ in the two decay channels presents a remarkable linear behavior with $\cos \theta_p$ in the range of  $|\cos\theta_p |\lesssim 0.75$. This surprisingly simple dependence on  $\cos \theta_p$ reflects the opposite behaviour of the helicity distributions, which is originated from the contribution of the final state interaction between the $p$ or $\bar p$ with $\pi^+$ or $K^+$. 

The key point to arrive at such a feature within the FBS framework is the dominance of the $p\bar p$,  $\pi^+p$ and $K^+\bar p$ interactions below invariant masses of 2\,GeV/c$^2$. This phenomenological information simplifies the plausible solution of the FBS equations.
We concluded that the difference between the  $B^+\to p\bar p\pi^+$ and $B^+\to p\bar pK^+$ normalized helicity distributions is almost linear in $cos\theta_p$ as indicated by the LHCb data. 

Our findings are promising and a motivation for further theoretical efforts to quantitatively solve the FBS equations in Minkowski space. 
This is a challenging task, but it could provide a framework for future studies of these decays, 
including the CP violation also observed by the LHCb collaboration in these channels.

\begin{acknowledgments}
This work was financed in part by the Funda\c c\~ao
de Amparo \`a Pesquisa do Estado de S\~ao Paulo (FAPESP),~Brazil, 
and Conselho Nacional de Desenvolvimento 
Cient\'ifico e Tecnol\'ogico~(CNPq), Brazil, 
This work was part of the project, 
Instituto Nacional de Ci\^{e}ncia e Tecnologia - Nuclear Physics and Applications 
(INCT-FNA), Brazil, No. 464898/2014-5, and FAPESP grant 
No. 2019/07767-1 (T.F.), and Coordena\c c\~ao de Aperfei\c coamento 
de Pessoal de N\'ivel Superior - Brazil (CAPES) - Finance Code 001.
\end{acknowledgments}

\newpage 

\appendix

\section{Faddeev-Bethe-Salpeter approach }
\label{app:FBS}

The full three-body T-matrix gives the final state interaction between the mesons in the three-body decay channel. It is a solution of the Bethe-Salpeter (BS)
equation, which will be written in the Faddeev form. We consider spinless particles,
disregard self-energies and three-body irreducible diagrams. Under these assumptions,
the interactions between the mesons are assumed to be only due to two-body interactions.
To be concise the momentum dependences will be omitted in the discussion below.

The three-particle BS equation for the T-matrix can be
written as
\begin{eqnarray}
T_{3\to3}=\sum V_{i} + \sum V_{i}G_0\,T , \label{eq1}
\end{eqnarray}
where the sum runs over the three two-body subsystems $i=(j,k)$. Formally, the potential
in the four-dimensional equation is built by multiplying the two-body interaction
$V^{(2)}_{jk}$ from all two-particle  irreducible diagrams in which particles \textit{j
}and \textit{k} interact, and by the inverse of the individual propagator of the spectator
particle $i$, $S_i$
\begin{eqnarray}
V=\sum_{i=1}^3 V_i\ ;~~V_i=V_{(2)jk}S^{-1}_i~.\label{presc}
\end{eqnarray}
 The propagator of particle $i$ is $S_{i}=\imath\left [ k_{i}^{2}-m_i^2+\imath
\epsilon\right]^{-1}$, $k_{i}$ being its four-momentum.
The three-particle free Green's function is
\begin{eqnarray}
G_0=S_{i}S_{j}S_{k}\, . \label{eq2}
\end{eqnarray}
Eq.~(\ref{eq1}) can now be rewritten in the
Faddeev form.  

The transition matrix is decomposed as
$$T_{3\to 3}=T^{(3)}_{1}+T^{(3)}_{2}+T^{(3)}_3$$ with the components $$T^3_i=V_{i}+V_{i}\,G_0\,T$$
The relativistic generalization of the connected Faddeev equations  is
\begin{eqnarray}
T^{(3)}_{i}=T_{jk}+T_{jk}G_0\left(T^{(3)}_{j}+T^{(3)}_{k}\right)
, \label{eq4}
\end{eqnarray}
where the two-body T-matrices are solutions of
\begin{eqnarray}
T_{jk}=V_{i}+V_{i}G_0T_{jk} , \label{eq5}
\end{eqnarray}
within the three-body system. The full $3 \to 3$ ladder scattering
series is summed up by solving the integral equations for the
Faddeev decomposition of the scattering matrix. Therefore, the three-body
unitarity holds for the 3$\to$3 transition matrix built from the solution
of the set of Faddeev equations (\ref{eq4}) below the threshold of particle production
from two-body collisions, where the two-body amplitude is unitary.

The full decay amplitude, Eq.~\eqref{eq:Ampl}, can be decomposed according to Eq.~\eqref{eq4}
as 
\begin{eqnarray}
{\mathcal A}=D+ \sum F_i \ , \label{d1}
\end{eqnarray}
where the Faddeev components of the decay vertex are
\begin{equation}
F_i=T_i^{(3)}\,G_0\,D \ . \label{d2}
\end{equation}
They are solutions of the connected equations
\begin{eqnarray}
F_{i}=F_{0,i}+T_{jk}G_0\left(F_{j}+F_{k}\right)
, \label{eq40}
\end{eqnarray}
with
\begin{eqnarray}
F_{0,i}= T_{jk}\,G_0\,D
. \label{eq41}
\end{eqnarray}
The Faddeev equations for the decay vertex, Eqs.~(\ref{eq40})-(\ref{eq41}) are general
once self-energies and three-body irreducible diagrams are disregarded.  In the
following they will be particularized to allow a separable form of the three-body
decay amplitude.

Substituting Eq.~\eqref{eq41} in \eqref{eq40} one has that:
\begin{eqnarray}
F_{i}=T_{jk}G_0\left(D+F_{j}+F_{k}\right)\,
, \label{eq42}
\end{eqnarray}
and therefore:
\begin{eqnarray}
{\mathcal A}=D+ \sum_i T_{jk}G_0\left(D+F_{j}+F_{k}\right)\ , \label{eq43}
\end{eqnarray}
where the sum is understood as taken from the cyclic permutations of $\{i,j,k\}$. The Faddeev-Bethe-Salpeter equations (FBS) formally written in Eq.~\eqref{eq42}, will be detailed in explicit form in Minkowski space in the next subsection.

In our particular case of $B^+\to p\bar p m^+$ with $m=K$ or $\pi$, we have:
\begin{equation}
    \begin{aligned}
{\mathcal A}^{B^+}_{p\bar p m^+}=&D+~ T_{p\bar p}\;G_0\;\left(F_{\bar p}
+F_p\right) \\ &+T_{mp}\,G_0\,\left(D+F_m+F_p\right)\\
&+T_{m\bar p}\,G_0\,\left(D +F_m+F_{\bar p}\right) \, ,
\end{aligned}
\label{eq44}
\end{equation}
where the $T_{p\bar p}G_0 D$ has been removed from the second term in the first line of the equation above to avoid double counting. This is necessary considering our hypothesis of the formation of the final state $p\bar pm$ ($A_{p\bar p}$ in Fig.~\ref{fig:3bodyrescatt}) through $xy\to p\bar p$  re-scattering from mesonic channels.

Up to two loops in the re-scattering series, we have that:
\begin{equation}
    \begin{aligned}
{\mathcal A}^{B^+}_{ p\bar p m^+}=~&D+T_{mp}\,G_0\,D +T_{m\bar p}\,G_0\,D\\ +~& T_{p\bar p}\;G_0\;\left( T_{mp}\,G_0\,D+T_{m\bar p}\,G_0\,D
\right) \\ +~&T_{mp}\,G_0\,T_{m\bar p}\,G_0\,D\\
+~&T_{m\bar p}\,G_0\,T_{mp}\,G_0\,D +\cdots \, ,
\end{aligned}
\label{eq44-1}
\end{equation}
which could distribute the effect of the meson-proton/antiproton resonances by subsequent $p\bar p$ re-scattering in the phase-space beyond the resonance region.

\section{Companion functions}
\label{app:fi}
We analyzed the FBS equations imposing that at low energies $\pi^+p$,  $K^+\bar p $ and $p\bar p$ are the dominant interactions. In the last case, there is a huge increase of the  $p\bar p$ elastic and inelastic cross-sections close to the threshold, reflected in the scattering amplitude. This simplify the FBS equations~\eqref{eq47} for the $\bar p p\pi^+$ channel, as:
\begin{equation}
\hspace{-.3cm} f_{\pi^+}(k_{\pi^+})=
\int \frac{d^4q_{\bar p} }{(2\pi )^4} S_{\bar p}(q_{\bar p})S_{ p}(q_p)\tau_{\pi^+ p}(s_{\pi^+ p})f_{\bar p}(q_{\bar p}) 
\,,
\label{eq47-1}
\end{equation}
where $q_p=P-k_{\pi^+}-q_{\bar p}$,
\begin{multline}
f_{\bar p}(k_{\bar p})=f_{0,\bar p}(k_{\bar p})
\\
+\int \frac{d^4q_{\pi^+} }{(2\pi )^4} S_{\pi}(q_{\pi^+})S_{  p}(q_{ p})\tau_{p\bar p}(s_{p\bar p})f_{\pi^+}(q_{\pi^+}) 
\,,
\label{eq47-3}
\end{multline}
with $q_{ p}=P-k_{\bar p}-q_{ \pi^+}$,  
and 
\begin{multline}
f_{p}(k_{p})=f_{0,p}(k_{p})
\\
+\int \frac{d^4q_{\pi^+} }{(2\pi )^4} S_{\pi}(q_{\pi^+})S_{ \bar p}(k_{\bar p})\tau_{p\bar p}(s_{p\bar p})f_{\pi^+}(q_{\pi^+}) \\
+\int \frac{d^4q_{\bar p} }{(2\pi )^4} S_{\bar p}(q_{\bar p})S_{ \pi^+}(k_{\pi^+})\tau_{\pi^+ p}(s_{\pi^+ p})f_{\bar p}(q_{\bar p}) 
\,,
\label{eq47-2}
\end{multline}
with $k_{\bar p}=P-k_{p}-q_{ \pi^+}$ and $k_{\pi^+}=P-k_p-q_{\bar p}$. It is reasonable to have that $f_{0,p}=f_{0,\bar p}$ and thus:
\begin{equation}
 f_{p}(k)=  f_{\bar p} ( k)+h_{\pi^+}( k)\,, \label{eqfpfpi}
\end{equation}
where
\begin{equation}
h_{\pi^+}(k_{p})=\int \frac{d^4q_{\bar p} }{(2\pi )^4} S_{\bar p}(q_{\bar p})S_{ \pi^+}(k_{\pi^+})\tau_{\pi^+ p}(s_{\pi^+ p})f_{\bar p}(q_{\bar p}) 
\,.
\label{eq477}
\end{equation}

Analogously for the $\bar p pK^+$ decay channel, the FBS equations become:
\begin{equation}
\hspace{-.3cm}f_{K^+}(k_{K^+})=\int \frac{d^4q_{ p} }{(2\pi )^4} S_{p}(q_{p})S_{ \bar p}(q_{\bar p})\tau_{K^+ \bar p}(s_{K^+ \bar p})f_{p}(q_{ p}) 
\,,
\label{eq47-4}
\end{equation}
with $q_{\bar p}=P-k_{K^+}-q_{ p}$, and
\begin{multline}
f_{p}(k_{ p})=f_{ 0,p}(k_{ p})
\\
+\int \frac{d^4q_{K^+} }{(2\pi )^4} S_{K}(q_{K^+})S_{\bar p}(q_{\bar p})\tau_{p\bar p}(s_{p\bar p})f_{K^+}(q_{K^+}) 
\,,
\label{eq47-6}
\end{multline}
where $q_{\bar p}=P-k_{ p}-q_{ K^+}$. The remaining companion amplitude is obtained as:
\begin{equation}
f_{\bar p}(k_{\bar p})=f_{p}(k_{\bar p})+h_{K^+}(k_{\bar p})\,, \label{eqfpfK}
\end{equation}
where
\begin{equation}
h_{K^+}(k_{\bar p})=\int \frac{d^4q_{p} }{(2\pi )^4} S_{p}(q_{ p})S_{ K^+}(k_{ K^+})\tau_{K^+ \bar p}(s_{K^+ \bar p})f_{ p}(q_{ p}) 
\,,
\label{eq47-5}
\end{equation}
with  $k_{ K^+}=P-k_{\bar p}-q_{ p}$. 

The amplitude $f_{\bar p}(k_{\bar p})$ is a sum of the amplitudes $f_{ p}(k_{\bar p})$ and $h_{K^+}(k_{\bar p})$ in Eq.~\eqref{eqfpfK}, whereas in $p\bar p\pi^+$ channel is the difference, as we observe in Eq.~\eqref{eqfpfpi}.
 This is a key difference between the amplitudes for the two decay channels.  

In the decays $B^+ \to p \bar{p} \pi^+$ and $B^+ \to p \bar{p} K^+$, the roles of the companion functions in Eqs.~\eqref{eqfpfpi} and \eqref{eqfpfK} are reversed.  Indeed, if in Eqs.~\eqref{eq47-1} and \eqref{eq47-3}, one exchanges $p\leftrightarrow {\bar p}$ and $\pi^+\to K^+$ one arrives to Eqs.~\eqref{eq47-4} and
\eqref{eq47-6}, respectively, and therefore the physical role played by the proton in the $B^+\to p\bar p \pi^+$ is played by the antiproton in the $B^+\to p\bar p K^+$ decay.

\section{Kinematics}
\label{app:kinematics}

The four-momenta for the three particles in the final state are  
\begin{equation}
    k_p, \quad k_{\bar p}, \quad k_m \quad k_B= k_p+ k_{\bar p}+ k_m\, ,
\end{equation}
and fulfill momentum conservation. The two-body subsystem masses are:
\begin{equation}
\begin{aligned}
  & M^2_{p\bar p}= (k_p+k_{\bar p})^2=m^2_p+m^2_{\bar p}+2 k_p\cdot k_{\bar p}, 
  \\   &M^2_{pm}= (k_p+k_{m})^2=m^2_p+m^2_m+2 k_p\cdot k_{m},
  \\
  &M^2_{\bar pm}= (k_m+k_{\bar p})^2=m^2_m+m^2_{\bar p}+2 k_m\cdot k_{\bar p},
  \\
 & M^2_B= (k_p+k_{\bar p})^2+m^2_m+ 2 (k_p+k_{\bar p})\cdot k_m
 \end{aligned}
 \label{eq:scalprod-1}
\end{equation}
Relation between the Dalitz variables:
\begin{equation}
\begin{aligned}
  M^2_B=& m^2_p+m^2_{\bar p}+m^2_m+ 2 k_p\cdot k_{\bar p}
 \\&+2 k_p\cdot k_m
 +2 k_m\cdot k_{\bar p}
 \\
 \\
M^2_B=& M^2_{p\bar p}+M^2_{pm}+M^2_{\bar pm}-2m^2_p-m^2_m
 \end{aligned}
 \label{eq:scalprod-2}
\end{equation}
other relations:
 \begin{equation}
\begin{aligned}
 & {k_p+k_{\bar p} \over M_{p\bar p}}\cdot k_m =
 {M^2_B-M_{p\bar p}^2-m^2_m \over 2\, M_{p\bar p}}
 \\
 & 
k_m\cdot k_{\bar p}=\frac12(M^2_{\bar pm}-m^2_m-m^2_{\bar p})\, . 
\end{aligned}
\end{equation}

The angle between the positively charged meson and antiproton in the rest frame of $p\bar p$ can be derived following the manipulations below:
\begin{equation}
    \begin{aligned}
&k_{p\bar p}=k_p+k_{\bar p}\, ,\\ &
k_{\bar p}\cdot k_m =
{M_{p\bar p} \over 2} k^0_m-\frac12
{\sqrt{M^2_{p\bar p}-4 m_p^2} }\, |\mathbf{k}_m|\, \cos \theta_p \, ,\\
& k^0_m= {k_p+k_{\bar p} \over M_{p\bar p}}\cdot k_m=\sqrt{m^2_m+|\mathbf{k}_m|^2}\, ,
  \end{aligned}
  \label{eq:costhetapaux}
\end{equation}
and we get:
\begin{equation}
    \begin{aligned}
& |\mathbf{k}^m_{p\bar p}|=\sqrt{\Bigg({k_p+k_{\bar p} \over M_{p\bar p}}\cdot k_m\Bigg)^2-m^2_m}\\
&=\sqrt{\Bigg({M^2_B-M_{p\bar p}^2-m^2_m \over 2\, M_{p\bar p}}\Bigg)^2-m^2_m}\, .
    \end{aligned}
    \label{eq:moduluskm}
\end{equation}

Finally, from Eqs.~\eqref{eq:costhetapaux} and \eqref{eq:moduluskm} the angle between the positively charged meson and the antiproton, $\theta_p$, can be obtained as:
\begin{equation}
    \begin{aligned}
 & \cos \theta_p= { (k_p+k_{\bar p}) \cdot k_m-2k_{\bar p}\cdot k_m  \over \sqrt{M^2_{p\bar p}-4 m_p^2}  |\mathbf{k}_m|}
\\
 &= {k_p \cdot k_m-k_{\bar p}\cdot k_m  \over \sqrt{M^2_{p\bar p}-4 m_p^2} \sqrt{\Bigg({M^2_B-M_{p\bar p}^2-m^2_m \over 2\, M_{p\bar p}}\Bigg)^2-m^2_m}} \, .
 \end{aligned}
\end{equation}
with further manipulations:
\begin{equation}
  \begin{aligned}
 &  \cos \theta_p= \\ &= {M^2_{pm}-M^2_{\bar p m} \over \sqrt{M^2_{p\bar p}-4 m_p^2} \sqrt{\Bigg({M^2_B-M_{p\bar p}^2-m^2_m \over \, M_{p\bar p}}\Bigg)^2-4m^2_m}} \, ,
  \end{aligned}
  \label{eq:costhetap}
\end{equation}
and the final expression in terms of $M_{p\bar p}$:
\begin{equation}
    \begin{aligned}
&  \cos \theta_p
=\\ & ={M^2_B-M_{p\bar p}^2 -2M^2_{\bar pm}+2m^2_{ p}+m^2_m \over  \sqrt{M^2_{p\bar p}-4 m_p^2} \sqrt{\Bigg({M^2_B-M_{p\bar p}^2-m^2_m \over \, M_{p\bar p}}\Bigg)^2-4m^2_m}} 
    \end{aligned}
    \label{eq:costhetap1}
\end{equation}
The mass of the pair $m\bar p $ as a function of  $\theta_p$ is:
\begin{equation}
\begin{aligned}
M^2_{\bar pm}&=\frac12\Big(M^2_B-M_{p\bar p}^2+m^2_m +2m^2_{p} \Big)
\\ &-\frac12\cos \theta_p \;\sqrt{M^2_{p\bar p}-4 m_p^2}
\\ &\times \sqrt{\Bigg({M^2_B-M_{p\bar p}^2-m^2_m \over \, M_{p\bar p}}\Bigg)^2-4m^2_m}
\,, 
    \end{aligned}
\end{equation}
from this formula, we find that for $\cos\theta_p=1$ the minimum value of $M_{\bar p m}$ is found for each given $M^2_{p\bar p}$.

\vspace*{2mm}

%\bibliographystyle{aps}
%\bibliography{references}
%
\end{document}